\documentclass[12pt]{article}

\usepackage{amsfonts,graphicx}

\begin{document}
\thispagestyle{empty}
\begin{flushright}
UCRHEP-T387\\ 
hep-ph/0504081\\
April 2005\
\end{flushright}
\vspace{0.5in}
\begin{center}
{\LARGE	\bf Hybrid Seesaw Neutrino Masses\\ with $A_4$ Family Symmetry\\}
\vspace{1.5in}
{\bf Shao-Long Chen, Michele Frigerio, and Ernest Ma\\}
\vspace{0.2in}
{\sl Physics Department, University of California, Riverside, 
California 92521\\}
\vspace{1.0in}
\end{center}

\begin{abstract}
We consider the scenario in which neutrino data are explained by the 
interplay of type I and II seesaw terms in the Majorana neutrino mass
matrix ${\cal M}_\nu = {\cal M}_L - {\cal M}_D {\cal M}_R^{-1} {\cal M}_D^T$. 
We construct a predictive model with ${\cal M}_L$ proportional to the unit 
matrix, 3 diagonal texture zeros in ${\cal M}_R$, and ${\cal M}_D$ diagonal. 
We show how this pattern can be maintained by the non-Abelian discrete 
symmetry $A_4$, and discuss its phenomenological consequences.
It turns out that the two types of seesaw give contributions 
of the same order to ${\cal M}_\nu$. 
In the CP conserving case, we find $\sin\theta_{13}\approx 2/(\tan 2\theta_{23}
\tan 2\theta_{12})$ and we predict 
inverted (normal) ordering of the mass spectrum
for $\tan^2\theta_{12} < 0.5 ~(>0.5)$.
\end{abstract}    

\begin{center}
\end{center}

\newpage
\baselineskip 24pt

In many well-motivated extensions of the Standard Model of particle 
interactions, the Majorana neutrino mass matrix is in general given by
\begin{equation}
{\cal M}_\nu = {\cal M}_L - {\cal M}_D {\cal M}_R^{-1} {\cal M}_D^T ~,
\label{master}\end{equation}
where the first term comes from the coupling of two left-handed 
neutrinos to a heavy Higgs triplet with a naturally small vacuum expectation 
value (type II seesaw \cite{seesawII}) and the second term comes from 
the canonical seesaw mechanism \cite{seesaw} assuming the existence of heavy 
singlet right-handed neutrinos.  In the past, perhaps for reasons of 
simplicity or economy, the common practice was to assume the dominance of 
one or the other of these two terms.  After all, if both terms were 
considered, predictability would be largely lost.  However, if there 
exists a symmetry which limits the forms of both terms, a simple and 
realistic Hybrid Seesaw model may still emerge.

An early discussion of the role of the two contributions in the generation 
of a large mixing angle can be found in \cite{wetterich}. 
One common strategy (see e.g. \cite{ak05} and references therein) was to 
assume a symmetry such as $SO(3)$ which requires ${\cal M}_L$ to be 
proportional to the unit matrix, but allow it to be broken arbitrarily in 
the second term.  Another recent paper \cite{RX} applies an $S_3\times S_3$ 
symmetry to both terms, but a number of symmetry-breaking parameters are 
needed to fit data.  Here we propose that the structure of both terms is 
fixed by the same family symmetry and thus obtain the first example of a 
predictive Hybrid Seesaw model with a well-defined symmetry (the discrete 
group $A_4$) for the complete Lagrangian. 

Consider first the type I contribution.  If ${\cal M}_D$ 
is diagonal (which may be maintained by the $A_4$ symmetry), then the 
texture zeros of ${\cal M}_R$ are reflected in ${\cal M}_\nu$ as zero 
sub-determinants \cite{m05}.  In fact, in the perspective of the type I 
seesaw formula, instead of the texture zeros of ${\cal M}_\nu$ \cite{fgm02}, 
those of ${\cal M}_R$ \cite{l04} are expected to have a deeper theoretical 
meaning (the two types of zeros may be related \cite{KKST}). 
In particular, consider the case
\begin{equation}
{\cal M}_R = \pmatrix{0 & \times & \times \cr \times & 0 & \times \cr \times 
& \times & 0},
\label{MR}\end{equation}
where $\times$ denotes a nonzero entry. This structure is rather unique, 
as it is the only possibility to have more than 2 zeros in ${\cal M}_R$ 
(and therefore less than 4 free parameters) without inducing 2 or more 
zeros in ${\cal M}_R^{-1}$.  Assuming ${\cal M}_D$ diagonal, one then obtains
\begin{equation}
{\cal M}^I_\nu = \frac 1a \pmatrix{a^2 & ab & ac \cr ab & b^2 & -bc \cr 
ac & -bc & c^2},
\label{typeI}\end{equation}
which has no texture zero but 3 zero sub-determinants.
This three-parameter structure, as we will show, cannot reproduce all 
present neutrino data \cite{data}.  On the other hand, it is possible that 
a significant contribution comes from ${\cal M}_L$ and, if it is proportional 
to the unit matrix, Eq.~(\ref{master}) becomes
\begin{equation}
{\cal M}_\nu = \pmatrix{d+a & b & c \cr b & d+b^2/a & -bc/a \cr 
c & -bc/a & d+c^2/a},
\label{genM}\end{equation}
which (i) turns out to fit all present data and (ii) may be stabilized by a 
simple family symmetry, as shown below. This model of Hybrid Seesaw, 
depending on 4 parameters, is the most minimal constructed so far.

To maintain the pattern of ${\cal M}_\nu$ in Eq.~(4), a suitable family 
symmetry is $A_4$, the discrete group of the even permutation of four 
objects.  It is also the symmetry group of the regular tetrahedron
(Plato's "fire" \cite{fire}), and has been applied to the 
neutrino mass matrix in a number of ways \cite{mr01,bmv03}. The irreducible 
representations of $A_4$ are ${\bf1}, {\bf 1'}, {\bf 1''}, {\bf 3}$.
The group multiplication rule \cite{mr01} is
\begin{equation}
{\bf 3} \times {\bf 3} = {\bf 1} + {\bf 1'} + 
{\bf 1''} + {\bf 3}_1 + {\bf 3}_2 ~,
\end{equation}
where $\psi_i,~\varphi_j \sim {\bf 3}$ implies
\begin{eqnarray}
{\bf 1} &=& \psi_1 \varphi_1 + \psi_2 \varphi_2 + \psi_3 \varphi_3~, \\ 
{\bf 1'} &=& \psi_1 \varphi_1 + \omega^2 \psi_2 \varphi_2 + 
\omega \psi_3 \varphi_3~, \\ 
{\bf 1''} &=& \psi_1 \varphi_1 + \omega \psi_2 \varphi_2 + 
\omega^2 \psi_3 \varphi_3~, \\ 
{\bf 3}_1 &=& (\psi_2 \varphi_3, \psi_3 \varphi_1, \psi_1 \varphi_2)~, \\ 
{\bf 3}_2 &=& (\psi_3 \varphi_2, \psi_1 \varphi_3, \psi_2 \varphi_1)~, 
\end{eqnarray}
with $\omega = e^{2 \pi i/3}$.  

Here we 
make the following assignment: the 3 families of leptons transform as a 
triplet,
\begin{equation}
(\nu_i,l_i), ~ l^c_i, ~ \nu^c_i \sim {\bf 3}~,
\end{equation}
and the scalar sector consists of 
three Higgs doublets $\Phi_i \sim {\bf 1}, {\bf 1'}, 
{\bf 1''}$, one Higgs triplet $\xi \sim {\bf 1}$, and three Higgs singlets 
$\Sigma_i \sim {\bf 3}$.
This implies that the Dirac mass matrices 
linking $l_i$ to $l^c_j$ (${\cal M}_l$) as well as 
$\nu_i$ to $\nu^c_j$ (${\cal M}_D$) are both diagonal, 
with 3 independent entries each. Explicitly,  
\begin{equation}
\left(\begin{array}{c} m_e \\ m_\mu \\ m_\tau \end{array}\right) =
\frac{1}{\sqrt{3}}\left(\begin{array}{ccc}
1 & 1 & 1 \\
1 & \omega & \omega^2 \\
1 & \omega^2 & \omega \\
\end{array}\right)
\left(\begin{array}{c} y_{l1} \langle\Phi_1\rangle \\
y_{l2} \langle\Phi_2\rangle \\ y_{l3} \langle\Phi_3\rangle \\
\end{array}\right)~,
\end{equation}
where the Yukawa couplings $y_{li}$ should be tuned to fit the charged lepton
masses, as in the Standard Model.
Our assignment also implies that ${\cal M}_L$ 
in Eq.(\ref{master}), which is generated by $\langle \xi \rangle$, is 
proportional to the unit matrix and ${\cal M}_R$ has nonzero off-diagonal 
entries, as in Eq.(\ref{MR}): $({\cal M}_R)_{ij}=f_R \langle \Sigma_k \rangle$ 
with $i\ne j\ne k$. Notice that, even if $\langle \Sigma_k \rangle$ were 
related among each other by the symmetry of the scalar potential, the 
parameters $a,b,c$ in Eq.(\ref{typeI}) would be completely independent, since 
they are determined by the 3 unknown diagonal entries of ${\cal M}_D$.
This is what is needed to obtain Eq.(\ref{genM}).

However, the bare Majorana mass term $\nu^c_i \nu^c_i$ is invariant 
under $A_4$ and it cannot be removed by hand.  
This is a generic issue in models with texture zeros in the mass matrix of 
gauge singlets.  Thus one is naturally led to consider a left-right 
gauge extension of the Standard Model with $(l^c,\nu^c)$ transforming as a 
doublet under $SU(2)_R$.  In that case, these bare mass terms are forbidden 
by the gauge symmetry and $\Sigma_i$ should now be considered as 
triplets under $SU(2)_R$, i.e. the counterpart of $\xi$ which is a triplet 
under $SU(2)_L$.  In this way our initial assumption in Eq.(\ref{MR}) is 
justified and 
the pattern of our proposed Hybrid Seesaw model in Eq.(\ref{genM}) 
is completely stabilized.

Let us briefly consider the phenomenology associated with the three $SU(2)_L$
doublets $\Phi_i$. Since charged lepton Yukawa couplings are diagonal, Lepton
Flavor Violation processes are suppressed by the smallness of neutrino 
masses and therefore negligible. The Standard Model like Higgs doublet is 
given by $\Phi = (v_1\Phi_1+v_2\Phi_2+v_3\Phi_3)/v$, where 
$v^2=v_1^2+v_2^2+v_3^2=(174$ GeV$)^2$. The orthogonal combinations $\Phi'$ and
$\Phi''$ decay into $e^+ e^-$, $\mu^+ \mu^-$
and $\tau^+ \tau^-$ with similar rates (the couplings are of the order 
$m_\tau/v$, the exact values depending on the scalar potential parameters).
One loop contributions to the anomalous magnetic moment of the muon, $g_\mu-2$,
are induced, but their size is generically negligible even for Higgs masses as 
light as 100 GeV (for a precise estimation in a similar model, see \cite{Z22}).

Notice also that, if the 3 families of quarks transform as an $A_4$ triplet
in the same way as leptons, up and down quark mass matrices are both diagonal,
thus describing in first approximation the smallness of CKM mixing angles.
Then, since all fermions transform in the same way under $A_4$, they may be 
embedded in multiplets of a Grand Unified gauge group. However, the 
construction of an appropriate scalar sector is highly non-trivial and beyond 
the purposes of the present paper.

Let us study the phenomenological implications
for neutrino masses and mixing angles.  
Data on neutrino oscillations \cite{data} indicate that $\theta_{23}$ is
close to maximal and $\theta_{13}$ is small. One can check that 
the matrix structure (\ref{genM}) may accommodate $\theta_{23}=\pi/4$ and 
$\theta_{13}=0$ if and only if $b^2 = c^2$. It is useful to discuss first this
limiting case and, in the following, possible deviations from it.

\noindent\underline{\bf Case $b=c$ :} The matrix ${\cal M}_\nu$ 
has a form \cite{m02} such that $\theta_{13} = 0$ and 
$\theta_{23} = \pi/4$.  In the basis spanning $\nu_e$, $(\nu_\mu + 
\nu_\tau)/\sqrt 2$, and $(\nu_\tau - \nu_\mu)/\sqrt 2$, it 
becomes
\begin{equation}
{\cal M}_\nu = \pmatrix{d+a & \sqrt 2 b & 0 \cr \sqrt 2 b & d & 0 \cr 
0 & 0 & d+2b^2/a}.
\label{2b2}\end{equation}
The parameters $a,b,d$ are in general complex. Diagonalizing ${\cal M}_\nu 
{\cal M}_\nu^\dagger$, we obtain
\begin{equation}
\tan 2 \theta_{12} = {2|B| \over |d|^2 - |d+a|^2}~,
\label{t12}\end{equation}
where $B = \sqrt 2 [2 Re(bd^*) + ab^*]$.  Notice that $\theta_{12}<\pi/4$ 
implies $|a|^2 + 2 Re(ad^*) < 0$. The mass squared differences are given by
\begin{equation}
\Delta m^2_{sol} \equiv |m_2|^2 - |m_1|^2 = 
\sqrt{(|d|^2-|d+a|^2)^2 + 4|B|^2} =
{|d|^2 - |d+a|^2 \over \cos 2 \theta_{12}}~,
\end{equation}
\begin{equation}
\pm \Delta m^2_{atm} \equiv |m_3|^2 - {1 \over 2} (|m_2|^2+|m_1|^2) = 
4 \left| {b^2 \over a} \right|^2 + 4 Re \left( {d^* b^2 \over a} \right) 
- 2|b|^2 + \frac 12 (|d|^2-|d+a|^2) ~.
\label{DMA}\end{equation}


\begin{figure}[tb]
\begin{center}
\includegraphics[width=270pt,height=220pt]{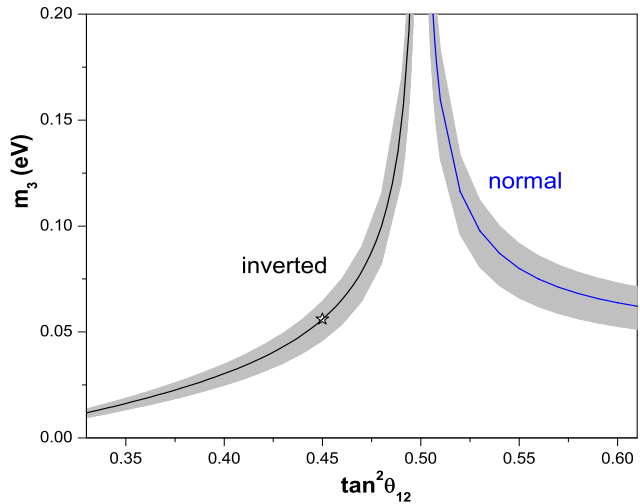}
\end{center}
\caption{The mass eigenvalue $m_3$ as a function of $\tan^2\theta_{12}$, in 
the limit $\theta_{13}=0$, $\theta_{23}=\pi/4$ and no complex phases 
(subcase (1)). The displayed interval is the $99\%$ C.L.
allowed range for $\tan^2\theta_{12}$ \cite{SV}.
The solid line corresponds to 
the best fit values for $\Delta m^2_{sol}$ and $\Delta m^2_{atm}$, whereas 
the shaded region accounts for the $99\%$ C.L. allowed ranges of
the mass squared differences. 
The star indicates the best fit.
The ordering of the mass spectrum is inverted for 
$\tan^2\theta_{12}<0.5$ (left branch) and normal for
$\tan^2\theta_{12}>0.5$ (right branch).}
\label{fig1}
\end{figure}

\noindent {\bf Subcase (1):}  If the parameters $a,~b$ and $d$ are real, 
they are 
uniquely determined by the experimental values of $\theta_{12}$, 
$\Delta m^2_{sol}$ and $\Delta m^2_{atm}$. In particular $\Delta m^2_{sol} 
\ll \Delta m^2_{atm}$ implies $2d\approx -a$, so that $d=0$ (pure type I 
seesaw) is not a solution, as already mentioned. Since
\begin{equation}
|m_3|^2 - {1 \over 2} (|m_2|^2+|m_1|^2) = - \frac{a^2 \tan^2 
2\theta_{12}}{2} \left(1-\frac 18 \tan^2 2\theta_{12}\right) + {1 \over 2} 
\Delta m^2_{sol} \left( 1 - {1 \over 2} \tan^2 2 \theta_{12} \right)~,
\label{phyDMA}\end{equation}
the ordering of the mass spectrum is inverted for $\tan^2\theta_{12}<0.5$, 
as favored (but only at about $1\sigma$ level) by present data. For the best 
fit values of oscillation parameters ($\tan^2 \theta_{12} = 0.45$, 
$\Delta m^2_{sol} = 8.0 \times 10^{-5}$ eV$^2$, $\Delta m^2_{atm} = 2.5 
\times 10^{-3}$ eV$^2$ \cite{SV}), 
one finds $a = \pm 0.057$ eV, $|b| = 0.049$ eV, 
$d = \mp 0.029$ eV and $|m_{1,2,3}|$ are respectively 0.0748, 0.0753, 
0.0560 eV, i.e. a mild inverted ordering. In this case the effective mass 
parameter relevant for neutrinoless $2\beta$-decay takes the value 
$m_{ee}\equiv |d+a| = 0.028$ eV. However, for values of $\tan^2 \theta_{12}$ 
closer to $0.5$, the absolute mass scale increases and the spectrum becomes 
quasi-degenerate, as shown in Fig.\ref{fig1}. Correspondingly, $m_{ee}
\approx \cos2\theta_{12} |m_1|$ becomes larger.

\noindent {\bf Subcase (2):} If the parameters $a$ and $d$ are real and
$b=i|b|$ is imaginary, then Eq.~(\ref{phyDMA}) is replaced by 
\begin{equation}
|m_3|^2 - {1 \over 2}(|m_2|^2+|m_1|^2) = {(\Delta m^2_{sol})^4 \sin^4 2 
\theta_{12} \over 16 a^6} + {(\Delta m^2_{sol})^3 \sin^2 2 \theta_{12} 
\cos 2 \theta_{12} \over 4 a^4} + {1 \over 2} \Delta m^2_{sol} \cos 2 
\theta_{12}.
\end{equation}
This is a solution with normal ordering and again the 3 experimental
conditions (best fit values) determine $a$, $|b|$, and $d$, i.e. 
$\pm0.0032$ eV, 0.0084 eV, and $\mp 0.0064$ eV, with $|m_{1,2,3}| = 0.011, 
0.014, 0.052$ eV respectively.  
Differently from subcase (1), the absolute mass scale depends weakly on
$\tan^2 \theta_{12}$ within the experimental range.


\begin{figure}[t]\begin{center}
\includegraphics[width=230pt]{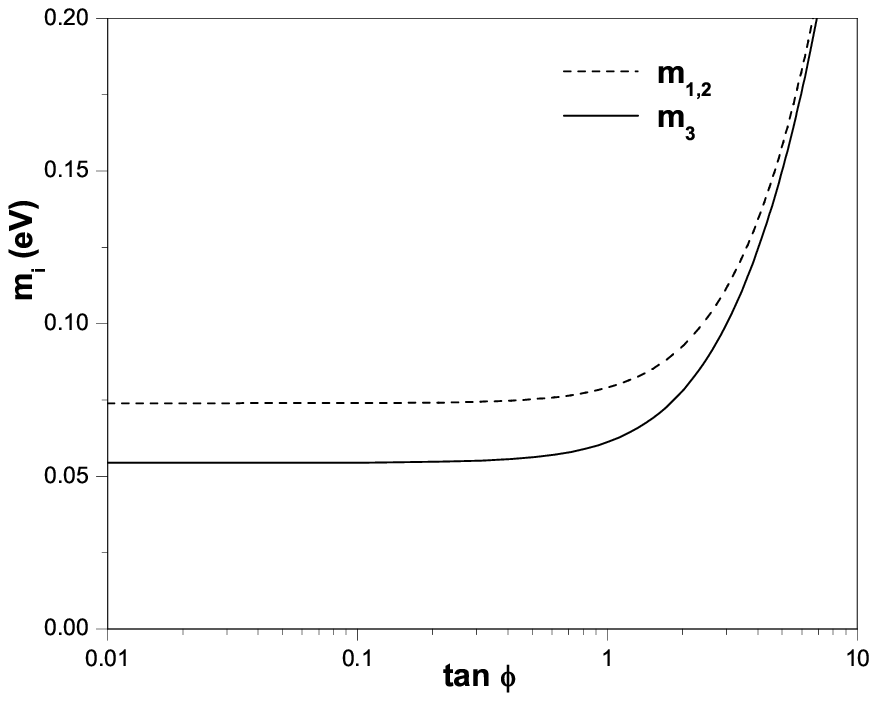}
\includegraphics[width=230pt]{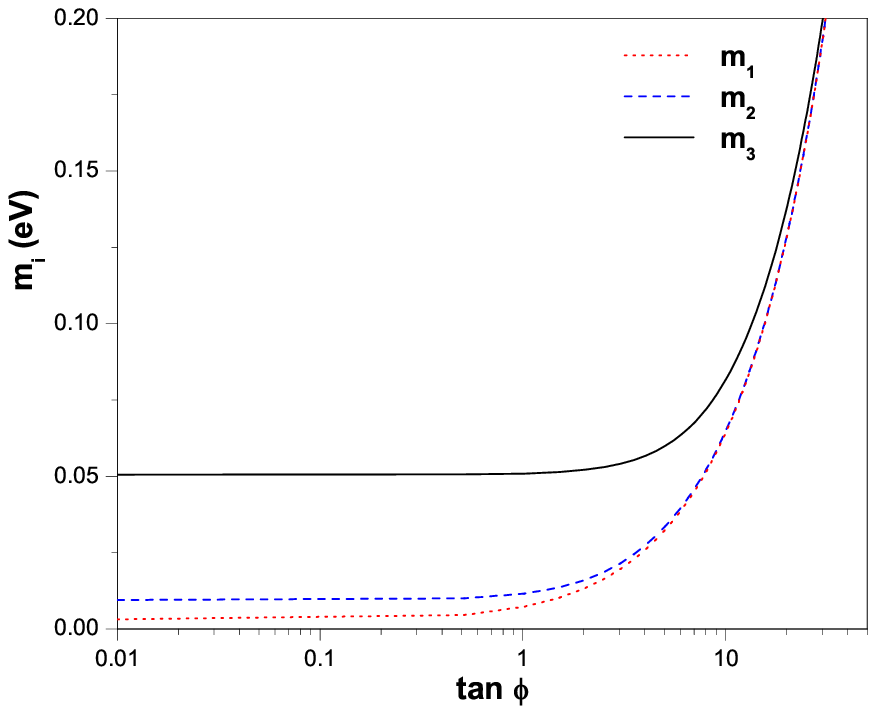}
\end{center}
\caption{The dependence of the neutrino mass eigenvalues $m_i$ 
on complex phases, in the
limit $\theta_{13}=0$ and $\theta_{23}=\pi/4$ (Eq.(\ref{2b2})). 
In the left panel we assume
$d$ real and $a,b$ having the same phase $\phi$ (subcase $(1')$). 
In the right panel we assume $d$ real, $a=|a|e^{i\phi}$ and $b=i|b|$ purely 
imaginary (subcase $(2')$). We take 
the best fit values of mass squared differences and $\tan^2\theta_{12}=0.45$.} 
\label{fig2}
\end{figure}

More in general, when $b=c$ there are two complex free parameters given by the
relative phases among $a$, $b$ and $d$. For illustration, let us
consider the following extensions of subcases (1) and (2), with only one 
additional degree of freedom. 
{\bf Subcase $(1')$:} Let $d$ be real with $a=|a|e^{i\phi}$ and 
$b=|b|e^{i\phi}$, i.e. $a$ and $b$ have the same phase. 
Then the 3 relations for $\tan 2 \theta_{12}$, 
$\Delta m^2_{sol}$, and $\Delta m^2_{atm}$ are exactly the same as in 
subcase (1), with the replacements $a\rightarrow |a|$, $b \rightarrow |b|$ and 
$d \rightarrow d \cos \phi$.  This means that we again have an 
inverted ordering for $\tan^2 \theta_{12} < 0.5$. Since only the
combination $d \cos \phi$ is determined by phenomenology and 
\begin{equation}
|m_3|^2 = \left| d + {2 b^2 \over a} \right|^2 = \left( d \cos \phi + 
2 \left| {b^2 \over a} \right| \right)^2 + (d \cos \phi)^2 \tan^2\phi ~,
\end{equation}
the overall scale of neutrino masses increases with increasing values of 
$\tan^2 \phi$. This means that the mass spectrum can be quasi-degenerate 
independently from the value of $\tan^2\theta_{12}$.
{\bf Subcase $(2')$:} Let $d$ be real with $a=|a|e^{i\phi}$ 
and $b=i|b|$. The 3 conditions are the same as in subcase (2),
with $a$ replaced by $|a|$
and $d$ by $d \cos \phi$.  We now have
\begin{equation}
|m_3|^2 = \left( d \cos \phi - 2 \left| {b^2 \over a} \right| \right)^2 
+ (d \cos \phi)^2 \tan^2\phi ~,
\end{equation}
so that, as in subcase $(1')$, 
the overall mass scale increases with  $\tan^2 \phi$.
The dependence of neutrino masses on $\phi$ is shown in Fig.\ref{fig2}, for 
both subcases $(1')$ and $(2')$.

\begin{figure}[p]
\begin{center}
\includegraphics[width=270pt,height=200pt]{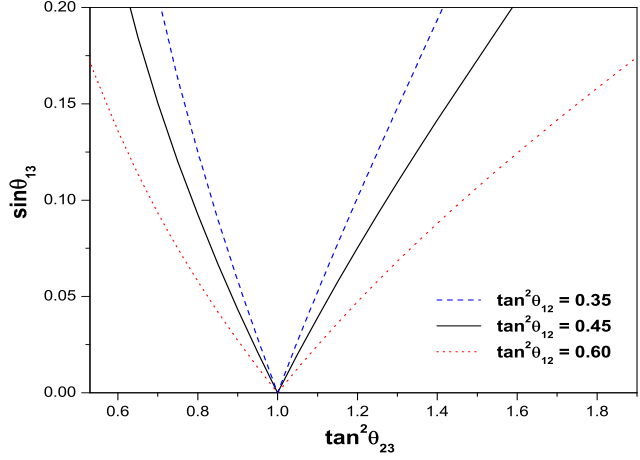}
\end{center}
\caption{The correlation between $\theta_{13}$ and $\theta_{23}$ in the
CP conserving case (no complex phases).
The displayed interval is the $99\%$ C.L.
allowed range for $\tan^2\theta_{23}$ \cite{SV}.
The curves depend only on the value of the solar mixing angle 
$\theta_{12}$ (they are independent of the neutrino mass spectrum).
}
\label{fig3}\end{figure}

\noindent\underline{\bf Case $b\ne c$ :} 
In this general case $\theta_{13}$ may be non-zero and 
$\theta_{23}$ may deviate from the maximal value $\pi/4$.
If one neglects complex phases, the type II term ${\cal M}_L = d {\mathbb I}$, 
being proportional to the identity, does not affect the mixing angles 
but only the mass spectrum: calling $\lambda_i$ the eigenvalues of 
${\cal M}^I_\nu$ in Eq.(\ref{typeI}), one has simply 
$m_i = d +\lambda_i$.
In order to extract the constraints on the mixing angles and $\lambda_i$, 
one should notice that $({\cal M}^I_\nu)^{-1}$ has 3 texture zeros on the 
diagonal, by construction.
It then follows that
\begin{equation}\begin{array}{l}
{\displaystyle 0=\frac{1}{\lambda_3}+\frac{1}{\lambda_2}+\frac{1}{\lambda_1}} 
~,\\ \\
{\displaystyle \tan^2\theta_{13} = \frac{\lambda_2\cos^2\theta_{12}+
\lambda_1\sin^2\theta_{12}}{\lambda_1+\lambda_2}} ~,\\ \\
{\displaystyle \tan2\theta_{23}=
\frac{\lambda_2\sin^2\theta_{12}+\lambda_1\cos^2\theta_{12}}
{(\lambda_1-\lambda_2)\cos\theta_{12}\sin\theta_{12}} ~
\frac{1}{\sin\theta_{13}}}~.\\
\end{array}
\label{tris}\end{equation}
Therefore, given the values of $\theta_{12}$ and 
$\theta_{23}$ ($\theta_{13}$), the ratio $\lambda_1/\lambda_2$ and 
$\theta_{13}$ ($\theta_{23}$) are predicted. 
In particular, taking into account that $\tan^2\theta_{13}<0.05\ll 1$, 
one finds
\begin{equation}
\sin\theta_{13} \approx \frac{1}{\tan 2\theta_{23}}\frac{2}{\tan 2\theta_{12}}
 ~,
\end{equation}
so that the size of the $1-3$ mixing angle is proportional 
to the deviation from maximal atmospheric mixing. 
This result is illustrated in Fig.\ref{fig3}, which shows that the 
present upper bound $\sin\theta_{13}< 0.2$ can be saturated, given the 
experimental uncertainty on $\theta_{23}$ and $\theta_{12}$.


\begin{figure}[p]
\begin{center}
\includegraphics[width=230pt,height=220pt]{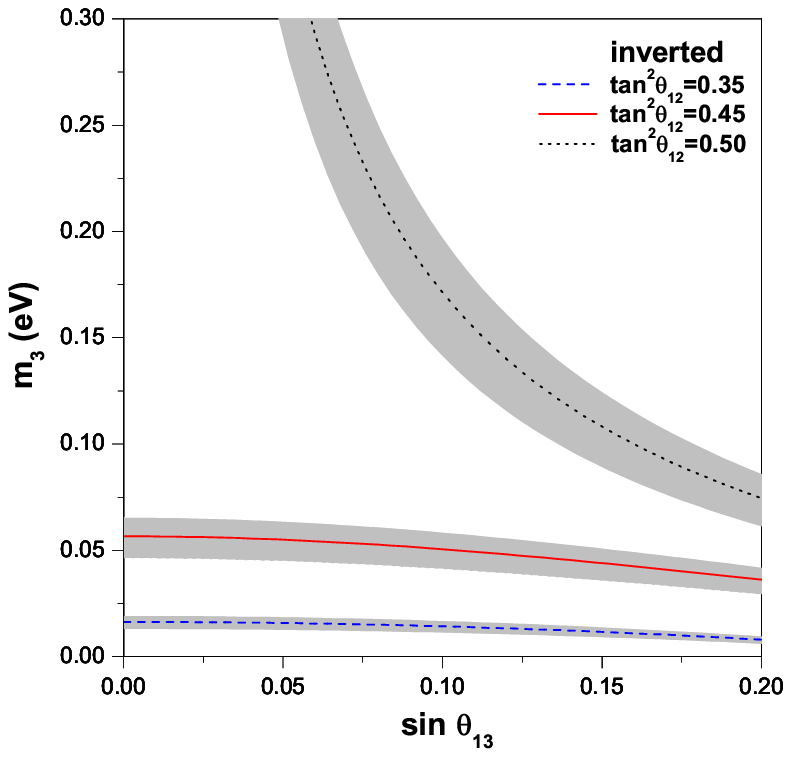}
\includegraphics[width=230pt,height=220pt]{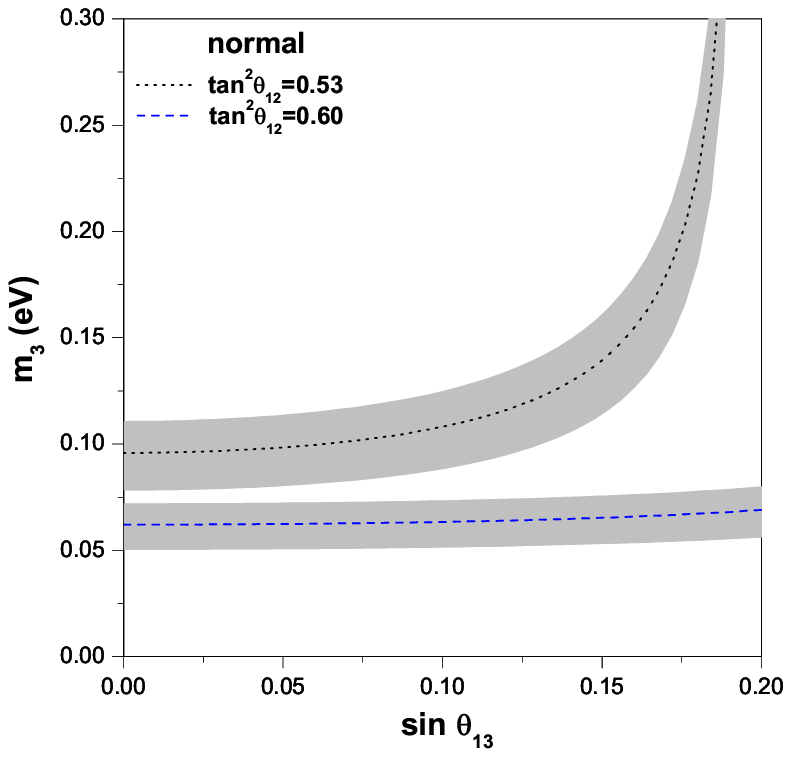}
\end{center}
\caption{The correlation between $m_3$ and $\theta_{13}$, for different 
values of the solar mixing angle $\theta_{12}$. The value of $\theta_{12}$ 
determines if the ordering of the mass spectrum is inverted (left panel) or 
normal (right panel).
The lines correspond to 
the best fit values for $\Delta m^2_{sol}$ and $\Delta m^2_{atm}$, whereas 
the shaded regions account for the $99\%$ C.L. allowed ranges of
the mass squared differences.
Complex phases are put to zero.}
\label{fig4}\end{figure}

\begin{figure}[tb]
\begin{center}
\includegraphics[width=230pt,height=220pt]{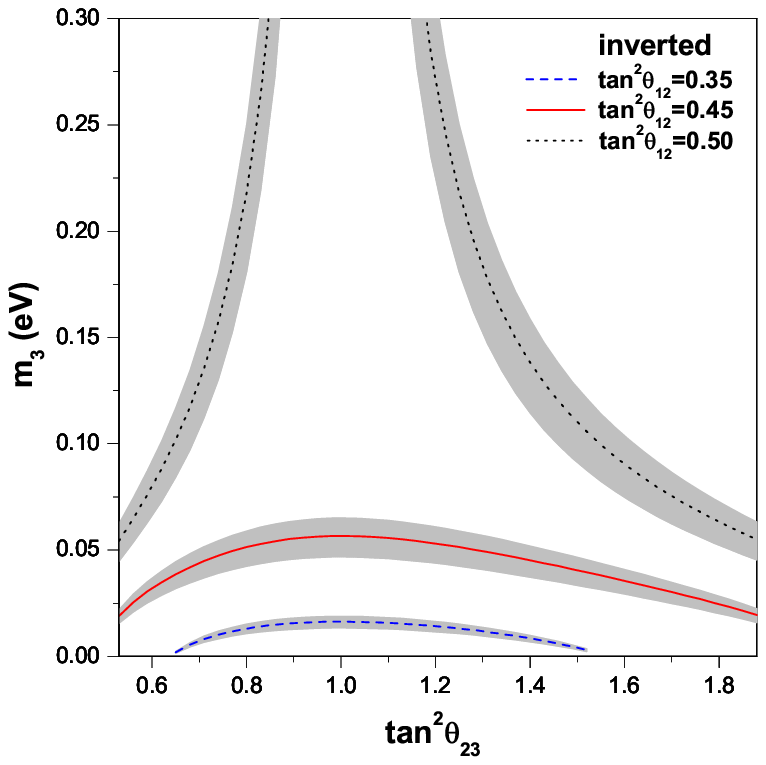}
\includegraphics[width=230pt,height=220pt]{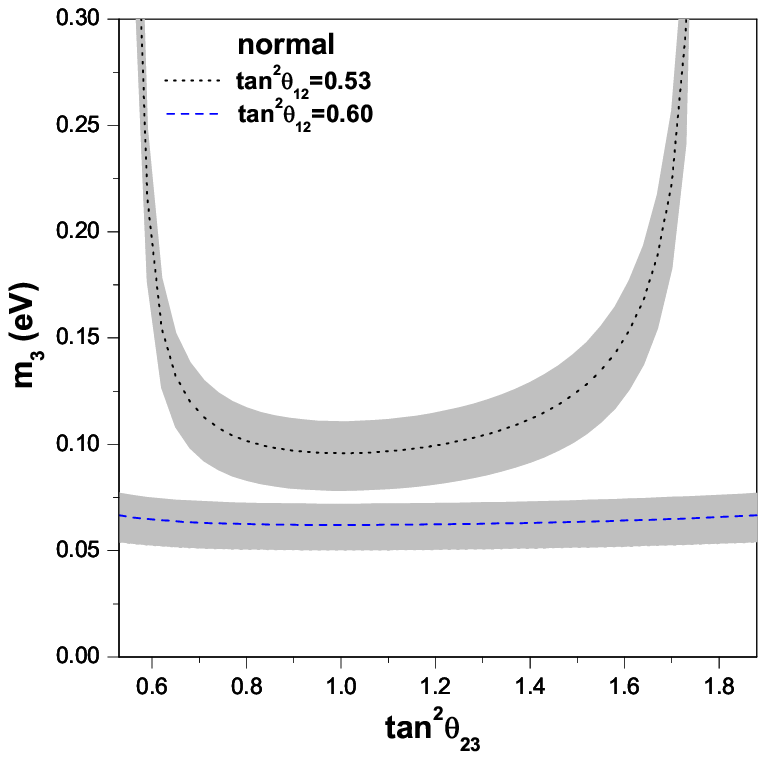}
\end{center}
\caption{The same as Fig.\ref{fig4} but as a function of $\theta_{23}$.
The displayed interval is the $99\%$ C.L.
allowed range for $\tan^2\theta_{23}$.
}
\label{fig5}\end{figure}

Since $m_i = d + \lambda_i$, the parameters
$d$ and $\lambda_{1,2}$ are uniquely determined once
$\Delta m^2_{sol}$ and $\Delta m^2_{atm}$ are given, 
so that the mass spectrum is predicted too. After some algebra, one finds that
the ordering is inverted for
\begin{equation}
\tan^2\theta_{12} < \frac {1+\tan^2\theta_{13}}{2-\tan^2\theta_{13}}
= 0.5 \div 0.54 ~
\label{perf}\end{equation}
and normal for $\tan^2\theta_{12}$ larger.
The absolute neutrino mass scale $|m_3|$
is shown, as a function of $\theta_{13}$
($\theta_{23}$), in Fig.\ref{fig4} (Fig.\ref{fig5}). 
The dependence of $|m_3|$ on $\theta_{12}$ is strong, 
analogously to the case $b=c$: quasi-degeneracy of the spectrum
(and accordingly sizable $m_{ee}$) 
is obtained for $\tan^2\theta_{12}$ close to the right-hand side of
Eq.(\ref{perf}).


If complex phases are introduced, the type II contribution will affect not 
only the mass spectrum but also the mixing angles. In general there will be 
more freedom to fit data and we do not elaborate further in this direction.
Just notice that $\theta_{13}\ne 0$ can be possibly associated with Dirac type
CP violation.

In conclusion, we have considered the Hybrid Seesaw scenario, where light
neutrino masses receive comparable contributions from super-heavy right-handed
neutrinos and scalar isotriplets.
We have shown that a family symmetry (the discrete group $A_4$) is able 
(i) to control the structure of type I and type II seesaw
terms at the same time and (ii) to restrict the number of free parameters 
so that predictions are possible and experimentally testable.

\noindent {\bf Acknowledgments:}
This work was supported in part by the U.~S.~Department of Energy
under Grant No.~DE-FG03-94ER40837.

\end{document}